\documentclass[prb,twocolumn,superscriptaddress,preprintnumbers]{revtex4-2}

\usepackage{graphicx}  % Include figure files
\usepackage{subfigure}
\usepackage{multirow}

\usepackage{fancyhdr}
\usepackage{longtable}
\usepackage[T1]{fontenc}
\usepackage{dcolumn}   % Align table columns on decimal point
\usepackage{dcolumn}   % Align table columns on decimal point

\usepackage{bm}        % bold math
\usepackage{amsfonts}  % Common math fonts
\usepackage{amsmath}   % Common math functions
\usepackage{amssymb}   % Common math symbols

\usepackage{soul}

\usepackage[colorlinks]{hyperref}% add hypertext
\hypersetup{
citecolor={blue},
urlcolor={blue}
}
\begin{document}

\title{Fractionalized Kohn-Sham Scheme for Strongly 
Correlated Electrons}
\author{Bo Zhao}
\affiliation{State Key Laboratory of Low Dimensional Quantum
Physics and Department of Physics, Tsinghua University, Beijing
100084, China}

\author{Jing-Yu Zhao}
\altaffiliation{Present address: Department of Physics and Astronomy, Johns Hopkins University, Baltimore, Maryland 21218, USA}
\affiliation{Institute for Advanced Study, Tsinghua University, Beijing
100084, China}

\author{Zheng Zhu}
\affiliation{Kavli Institute for Theoretical Sciences, University of Chinese Academy of Sciences, Beijing 100190, China}

\author{Jian Wu}
\affiliation{State Key Laboratory of Low Dimensional Quantum
Physics and Department of Physics, Tsinghua University, Beijing
100084, China}
\affiliation{Frontier Science Center for Quantum Information, Beijing, China.}

\author{Zheng Liu}
\email{zliu23@buaa.edu.cn}
\affiliation{School of Physics, Beihang University, Beijing
100191, China}

\begin{abstract}
We propose to expand the territory of density functional theory to strongly correlated electrons by reformulating the Kohn-Sham scheme in the representation of fractionalized particles. We call it the ``KS$^*$ scheme.'' Using inhomogeneous $t$-$J$ chains as a test bed, we show that the KS$^*$ scheme with simple local density approximation is able to achieve accurate ground-state energy and density distribution comparable to the density matrix renormalization group method, while the computational complexity is much lower.

\vspace{1.0em} % 调整间距
\noindent DOI: \href{https://doi.org/10.1103/PhysRevLett.134.136505}{10.1103/PhysRevLett.134.136505}%\doi{10.1103/PhysRevLett.134.136505} 
\end{abstract}

\maketitle
The success of density functional theory (DFT)~\cite{HK,DFT-review} for first-principles electronic structure calculations is built upon the Kohn-Sham (KS) scheme~\cite{KS}, which asserts that a noninteracting auxiliary system with a properly chosen exchange-correlation (XC) potential has the same ground-state (GS) density as the original interacting electrons. Together with the local density approximation (LDA) or its improved versions to the XC potential, the KS approach is proved to work well for a wide range of condensed matters and molecular systems~\cite{KohnReview}. However, it is also known to fail in systems dominated by electron-electron interactions~\cite{strong-correlation-DFT-1,strong-correlation-DFT-2}. The strong quantum correlation makes such systems lack resemblance to a noninteracting electron gas.

Tremendous efforts have been made to expand the territory of first-principles calculation to the strong-correlation regime, either by improving the XC potential~\cite{ODDFT, perdew2001jacob, SCAN, DeepMind} or hybridizing DFT with other many-body techniques~\cite{DMFTReview, GWDMFT, comdmftv2,GWreview}. In this Letter, we aim to explore a new direction by highlighting one prominent feature of strong correlation, namely, the emergence of exotic quasiparticles as a fraction of the electrons (for a recent overview, see Ref.~\cite{SachdevQPM} and references therein). Instead of further upgrading interaction corrections within the standard KS setup, we propose to construct new types of noninteracting auxiliary systems consisting of fractionalized particles rather than electrons, which we term as the ``{fractionalized KS (KS$^*$) scheme}''. 

Theoretically, the power of fractionalization lies in the possibility to transform strongly correlated electrons into weakly correlated quasiparticles. A well-known example is the spin-charge separation in a Luttinger liquid~\cite{LuttingerReview}, and more and more strongly correlated systems are becoming describable within this framework (see, e.g.,\ Parts II and IV in~\cite{SachdevQPM}). The expectation is that a well-chosen fractionalization scheme will largely reduce the complexity of the XC potential, such that satisfactory results could be achieved at the simple and efficient LDA level.

The general concept the of KS$^*$ scheme opens new possibilities to tackle different strongly correlated phases within DFT, but the design of the fractionalized auxiliary system and the density function(al)s requires theoretical thoughts and numerical tests in a case-by-case way. To implement this idea in an explicit strongly correlated system, we choose the 1D $t$-$J$ model as the first example. We write the Hamiltonian as
\begin{eqnarray}
 \hat H=\hat H_0+\hat H_{\mathrm{int}},
\end{eqnarray}
with
\begin{eqnarray}\label{H}
\hat H_0&=&-t\sum_{i\sigma}\left(c_{i\sigma}^{\dag}c_{i+1\sigma} + H.c.\right)\color{black}+\sum_iV_i\hat n_i \nonumber \\
\hat H_{\mathrm{int}}&=&J\sum_{i} \left(\mathbf{\hat S}_i \cdot \mathbf{\hat S}_{i+1}-\frac{\hat n_i\hat n_{i+1}}{4}\right).
\end{eqnarray}
The operator $c_{i\sigma}^\dag$ creates an electron at site $i$ with spin $\sigma$, the vector operator $\mathbf{\hat S_i}=\frac{1}{2}c_{i,\sigma}^{\dag}   (\hat{\bm{\sigma}}_{\sigma,\sigma'})c_{i,\sigma'}$ refers to the electron spin at site i with the Pauli matrices $\hat{\bm{\sigma}}=\{\hat{\sigma}_x,\hat{\sigma}_y,\hat{\sigma}_z \} $, $\hat n_i=\sum_{\sigma}c_{i\sigma}^{\dag}c_{i\sigma}$ is the electron number operator, and $V_i$ is the external potential.

The $t$-$J$ model arises from a downfold of the Hubbard model at the large-$U$ limit, which is thus  constrained by the no-double-occupancy condition
\begin{equation}
n_i \leq 1.
\end{equation}
%Nontrivial physics appears when the system is doped from the $n_i\equiv1$ limit. %The hole number at a site is defined as $1-n_i$.
Properly treating this constraint is essential to understand doped Mott insulators~\cite{DopeMott}. 

%In parallel with the full Hamiltonian of interacting electrons, $\hat H_0$ can be regarded as the discretized version of electron kinetic energy plus the lattice potential. $\hat H_{int}$ replaces the Coulomb interaction with an effective nearest neighbor form - when one of the two sites is empty, it is zero; when both sites are singly occupied, the two electrons are correlated by the antiferromagnetic Heisenberg superexchange. 

We have several motivations to take a detour from the full Hamiltonian of interacting electrons to such an effective lattice model. First, the strongly correlated regime of interacting electrons is typically accessed under a strong lattice potential, e.g.,\ in $d$,$f$-electron materials, where the lattice model represents a more efficient description than the plane wave continuum model~\cite{capelle2013density}. Second, while the more familiar Hubbard model still has a perturbative regime when the on-site interaction $U$ is small, the $t$-$J$ model focuses on the strong-correlation regime, as explicitly reflected by the no-double-occupancy constraint~\cite{DopeMott}. Third, despite the long-lasting challenge toward a thorough understanding of the 2D $t$-$J$ model and its implication on high-$T_c$ cuprate superconductors~\cite{DopeMott}, the Luttinger liquid behavior in 1D is quantitatively solvable, both analytically ~\cite{LiebWu} and numerically via the density matrix renormalization group (DMRG) method~\cite{DMRG}. Therefore, the performance of our new scheme can be systematically tested. For a lattice model, the in-principle existence of a universal density function of the total energy can be most conveniently proved by following L\'{e}vy and Lieb's two-step minimization procedure~\cite{levy1979,lieb1983}. In practice, the lattice DFT has been studied in a variety of setups~\cite{chayes1985density,penz2024geometrical,maoPRB,xuliming}. Once encouraging results have been obtained from this transparent effective model, we can move on to more complicated setups and ultimately toward realistic materials. 

Before invoking fractionalization, let us first revisit the standard KS scheme. For the discrete lattice model, we compact the density distribution as a vector: $\vec n \triangleq (n_1, n_2, n_3, ...)$. The KS scheme can be rephrased as follows:

KS: Once $t$ and $J$ are fixed in Eq.(\ref{H}), a so-called XC energy function $E_{\mathrm{xc}}(\vec n)$ \textit{in principle}  exists, such that self-consistently solving the auxiliary Hamiltonian:
\begin{eqnarray}\label{HKS}
\hat H_{\mathrm{KS}}=\hat H_0+\hat V_{\mathrm{xc}}
\end{eqnarray}
with 
\begin{eqnarray}\label{Vxc}
\hat V_{\mathrm{xc}}=\sum_i \frac{\partial E_{\mathrm{xc}}}{\partial n_i}\hat n_i
\end{eqnarray}
gives the exact GS density. The GS expectation value of $\hat H_0$ of the auxiliary system plus $E_{\mathrm{xc}}$ gives the exact GS energy.

%Applying DFT to this model equals to replacing $\hat H_{int}$ with a so-called exchange-correlation (xc) energy $E_{xc}$. Once $\hat H_{int}$ is fixed, DFT dictates that $E_{xc}$ is a function of $\vec n$.

LDA renders an approximated but straightforward construction of $E_{\mathrm{xc}}(\vec n)$ by referring to the XC energy per electron $\epsilon_{\mathrm{xc}}(n)$ of the homogeneous chain, i.e.,\ $V_i\equiv 0$ and $n_i=n$. The LDA formula is
\begin{eqnarray}\label{LDA}
    E_{\mathrm{xc}}(\vec n)\approx \sum_i n_i\epsilon_{\mathrm{xc}}(n_i).
\end{eqnarray}
Extending the formula to spin-polarized cases is straightforward, but we will focus on the nonmagnetic regime in the present Letter. 

\begin{figure}
    \centering
    \includegraphics[width=0.5\textwidth]{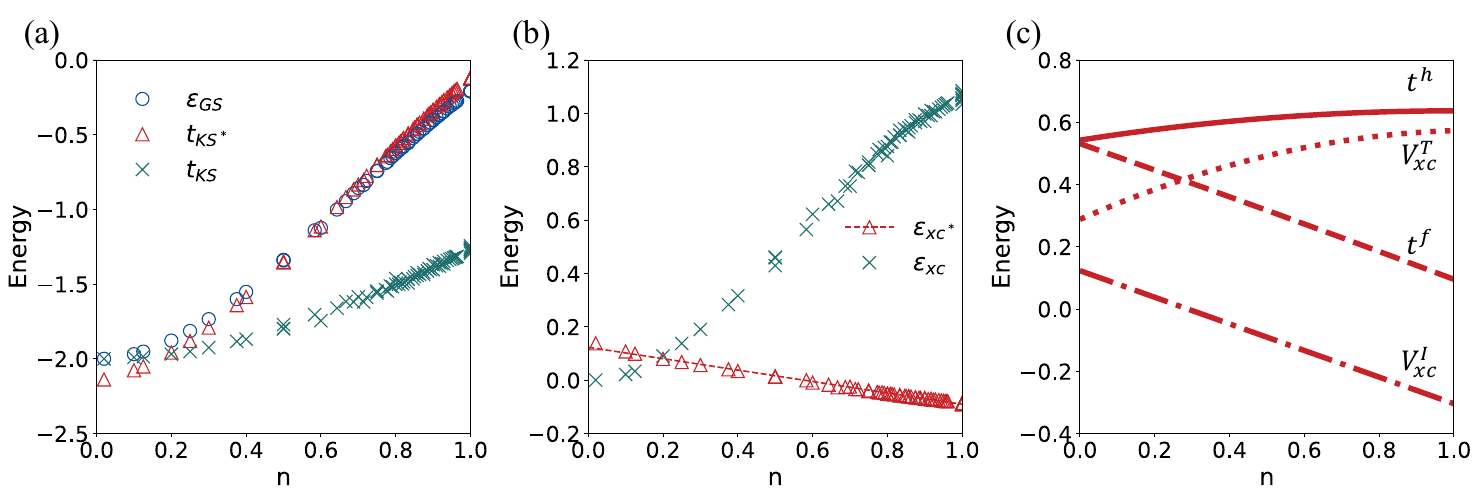}
    \caption{Homogeneous density dependence of (a) $\epsilon_{\mathrm{GS}}$, $t_{\mathrm{KS}}$, $t_{\mathrm{KS}^*}$; (b) $\epsilon_{\mathrm{xc}}$, $\epsilon_{\mathrm{xc}^*}$; and (c) $t^h$, $t^f$, $V_{\mathrm{xc}}^T$, $V_{\mathrm{xc}}^I$. See context for the definitions. Data shown in (a) and (b) are collected from calculations on finite $N_s$-site chains ($N_s=10\sim50$) containing different number of holes under the periodic boundary condition (PBC). $t_{\mathrm{KS}^*}$ in (a) is obtained by using the LDA parameters $A=0.45$, $B=-0.05$. The dashed line in (b) is the linear fitting $\epsilon_{\mathrm{xc}^*}=-0.21n_i+0.12$ and $\epsilon_{\mathrm{xc}}=1.22n_i-0.12$}.
    \label{fig:1}
\end{figure}

The required information to wrap up the computation is then reduced to $\epsilon_{\mathrm{xc}}(n)$ of the homogeneous chain. Like the role the quantum Monte Carlo method played for the homogeneous electron gas~\cite{LDACA}, DMRG can serve as a benchmark of the total energy per electron of the homogeneous chain [$\epsilon_{\mathrm{GS}}$, blue circles in Fig.\ \ref{fig:1}(a)] because DMRG is considered to be almost exact for 1D systems~\cite{DMRG,DMRG1993,DMRG2011}. The kinetic energy per electron of the noninteracting chain [$t_{\mathrm{KS}}$, green crosses in Fig.\ \ref{fig:1}(a)] can be easily calculated. By definition, $\epsilon_{\mathrm{xc}}=\epsilon_{\mathrm{GS}}-t_{\mathrm{KS}}$ [green crosses in Fig.\ \ref{fig:1}(b)]. 

For all the numerical calculations, we treat $t$ as the energy unit, and $J$ is fixed to be 0.3. We note that $J/t=0.3$ is the common ratio employed for numerical study of $t$-$J$ models~\cite{phase-diagram-1DtJ,NPB16Zhu,phase-diagram-1DtJ-pairhoppind,PRX22Weng} due to connections to the realistic parameters of cuprates (see, e.g.,\ Sec.\ II of~\cite{DopeMott}). DMRG computational details are given in Supplemental Material~\cite{SM}. 

One quick observation is that $t_{\mathrm{KS}}$ becomes significantly lower than $\epsilon_{\mathrm{GS}}$ when $n$ increases, reflecting that the antiferromagnetic superexchange frustrates the motion of electrons. The positive $\epsilon_{\mathrm{xc}}$ and negative $t_{\mathrm{KS}}$ largely cancel each other as $n$ approaches 1. We have tested to solve Eqs.\ (\ref{HKS})-(\ref{LDA}) self-consistently for several inhomogeneous density profiles. The performance is poor (Fig.\ S1~\cite{SM}). It also requires additional advanced techniques, e.g.,\ Gutzwiller projection, to take into account the no-double-occupancy constraint within this framework~\cite{PRB09Gutzwiller, JPC22Gutzwiller}. All these issues make it cumbersome to proceed along the conventional KS scheme.

We now switch to the KS$^*$ scheme. The starting point is that the general Luttinger liquid behavior of 1D interacting electrons asserts spin-charge separation. Accordingly, we assume a fractionalized auxiliary system consisting of (i) spinless particles each carrying a positive elementary charge (holons) and (ii) neutral spin-1/2 particles (spinons). The holons and spinons  move at their own velocities. We write their hopping amplitudes as $t_i^h$ and $t_i^f$, respectively, which typically differ from $t$ in $\hat H_{0}$, due to the renormalization effect of $\hat H_{\mathrm{int}}$. The external potential $V_i$ couples to the charge degree of freedom, i.e.,\ holons. Additionally, we add an XC potential $V_{\mathrm{xc}}$ as in the conventional KS scheme. $\hat H_{\mathrm{int}}$ manifests in $V_{\mathrm{xc}}$, as well as $t_i^h$ and $t_i^f$. The three quantities are considered to be functions of the density distribution. The no-double-occupancy constraint means that each site has three configurations: (i) empty, (ii) occupied by a single spin-up electron, and (iii) occupied by a single spin-down electron. Converting to the hole representation with $n_i=1$ as the charge neutral point, this is equivalent to each site being occupied either by a holon or a spinon. Note that hole doping is measured from the half filling $n_i=1$ limit, i.e., $N_h$ holes correspond to a total electron number $N_s-N_h$, where $N_s$ is the number of sites. Therefore, the total number of holons and spinons at each site should always be 1, which conveniently transforms the inequality constraint into the condition $
\langle h_i^{\dag}h_i+\sum_{\sigma}f_{i\sigma}^{\dag}f_{i\sigma}\rangle=1$, which can be enforced by introducing Lagrange multipliers $\lambda_i$, and by self-consistently determining $\lambda_i$ as a local chemical potential. Formally, we write a noninteracting Hamiltonian as follows:

\begin{eqnarray}\label{HKSS}
    \hat H_{\mathrm{KS}^*}=&-&\sum_{i}t^h_i(\vec n)h_{i}^{\dag}h_{i+1} -\sum_{i\sigma}t^f_i(\vec n)f_{i\sigma}^{\dag}f_{i+1\sigma}+H.c. \nonumber \\
    &+&\sum_i[V_i+V_{\mathrm{xc},i}(\vec n)](1-h_i^\dag h_i) \nonumber \\
    &+&\sum_i \lambda_i\left(h_i^\dag h_i+\sum_\sigma f_{i\sigma}^\dag f_{i\sigma}-1\right), 
\end{eqnarray}
in which \(h_i^{\dag}\) (\(f_{i\sigma}^\dag\)) creates a holon (spinon). %; $t_i^{h(f)}$ is the local hopping amplitude of the corresponding particle; and $V_{xc}$ is an effective XC potential. 
Note that $1-h_i^\dag h_i=\hat n_i$ within the no-double-occupancy subspace, consistent with our definition of hole doping concentrations. To keep the fermionic statistics of electrons, we assign \(f_{i\sigma}\) as a fermion operator and \(h_i\) as a hard-core boson operator. In practical calculations,  \(h_i\) will be transformed into a fermion operator via the 1D Jordan-Wigner transformation (see Sec.\ IV of Supplemental Material~\cite{SM}).

Based on Eq.\ (\ref{HKSS}), we propose the KS$^*$ scheme:

KS$^*$: Once $t$ and $J$ are fixed in Eq.\ (\ref{H}), function forms of  $t_i^{h(f)}(\vec n)$ and $V_{\mathrm{xc},i}(\vec n)$ {in principle} exist, such that solving Eq.\ (\ref{HKSS})
gives the exact GS density. The exact GS energy can be obtained from a corresponding energy function, which links to Eq.\ (\ref{HKSS}) via the variational principle.

%Generally speaking, we propose that the choice of a suitable KS auxiliary system for strongly-correlated electrons should be guided by physical intuitions. 

The proposal is intuitively backed by the Luttinger liquid behavior of 1D interacting electrons~\cite{HaldanLuttinger}. Most relevantly, an elegant slave-particle construction $c_{i\sigma}=h_i^{\dag}f_{i\sigma}$ is known (see, e.g.,\ Sec.\ XIII of~\cite{DopeMott}) to transform Eq.\ (\ref{H}) into:
\begin{eqnarray}\label{H_slave}
\hat H_0&=&-t\sum_{i\sigma}h_{i}^{\dag}h_{i+1}f_{i+1\sigma}^{\dag}f_{i\sigma} + H.c.+\sum_iV_i\hat n_i \nonumber \\
\hat H_{\mathrm{int}}&=&-\frac{J}{2}\sum_{i \sigma\sigma'} f_{i\sigma}^{\dag}f_{i+1\sigma} f_{i+1\sigma'}^{\dag}f_{i\sigma'}
\end{eqnarray}

Then, a mean-field(MF) treatment leads to the following noninteracting Hamiltonian~\cite{Phasestring1,Phasestring2}:
\begin{eqnarray}\label{MFS}
\hat H_{\mathrm{MF}^*}=&-&t\sum_{i\sigma}\langle f_{i+1\sigma}^{\dag}f_{i\sigma} \rangle h_{i}^{\dag}h_{i+1}\nonumber \\
&-&\sum_{i\sigma\sigma'}\left(\frac{J}{2}\langle f_{i+1\sigma'}^{\dag} f_{i\sigma'}\rangle+t \langle h_{i+1}^{\dag}h_{i} \rangle\right)f_{i\sigma}^{\dag}f_{i+1\sigma} \nonumber \\
&+&H.c.+\sum_iV_i\hat n_i
\end{eqnarray}

The relation between $\hat H_{\mathrm{MF}^*}$ and $\hat H_{\mathrm{KS}^*}$ is analogous to that between the Hartree-Fock mean-field treatment of the electron gas and the KS equations. For the electron gas, it is well known that the KS scheme not only renders a formal exactification, but also in many cases practically surpasses the mean field in accuracy~\cite{KohnReview}. We will see that the KS$^*$ scheme retains these advantages. 

$\hat H_{\mathrm{KS}^*}$ contains three density functions. In addition to the XC potential $V_{\mathrm{xc}^*}$, we introduce density dependence to $t_i^{h(f)}$. Physically, it means that $\hat H_{\mathrm{int}}$ also manifests in the motion of the fractionalized particles, as it should do. According to $\hat H_{\mathrm{MF}^*}$,
\begin{eqnarray}\label{eq:thtf}
    t_i^h&=& t\sum_\sigma \langle f_{i+1\sigma}^{\dag} f_{i\sigma}\rangle \nonumber \\
    t_i^f&=&\frac{J}{2}\sum_{\sigma'}\langle f_{i+1\sigma'}^{\dag} f_{i\sigma'}\rangle+t \langle h_{i+1}^{\dag}h_{i} \rangle.
\end{eqnarray}
However, we find that directly applying this formula tends to result in unrealistic oscillating $t_i^{h(f)}$ when $n_i$ is inhomogeneous. The failure can be most easily seen in a half filled four-site chain with the open boundary condition (OBC), for which $\hat H_{\mathrm{MF}^*}$ predicts an incorrect valence-bond-solid state (see Sec.\ II of Supplemental Material~\cite{SM}). Numerical comparisons between DMRG and $\hat H_{\mathrm{MF}^*}$ GSs on several inhomogeneous density profiles also indicate noticeable errors (Fig.\ S2~\cite{SM}).

Instead, we invoke LDA to $\langle h_{i+1}^{\dag}h_{i}\rangle$ and $\langle f_{{i+1}\sigma}^{\dag} f_{i\sigma}\rangle$ and require the density functions recover the behavior of the mean-field definition at the homogeneous limit (see Sec.\ III of Supplemental Material~\cite{SM}) in two aspects:(i) when $n_i=1$, $\langle h_{i+1}^{\dag}h_{i}\rangle=0$, $\langle f_{i+1\sigma}^{\dag}f_{i\sigma}\rangle=\frac{1}{\pi}$; and (ii) when $n_i$ approaches 1 from below, $\delta \langle h_{i+1}^{\dag}h_{i}\rangle\propto1-n_i$, $\delta\langle f_{i+1\sigma}^{\dag}f_{i\sigma}\rangle\propto (1-n_i)^2$. These constraints lead to
\begin{eqnarray}\label{eq:hhff}
    \langle h_{i+1}^{\dag}h_{i} \rangle_{\mathrm{LDA}} &\approx& A(1-n_i),  \nonumber \\ 
\langle f_{i+1\sigma}^{\dag} f_{i\sigma}\rangle_{\mathrm{LDA}} &\approx& \frac{1}{\pi}+B(1-n_i)^2,
\end{eqnarray}
in which $A$ and $B$ are two LDA parameters. Substituting Eq.\ (\ref{eq:hhff}) back into Eq.\ (\ref{eq:thtf}) decides the LDA forms of $t_i^{h(f)}$. %Hereafter, whenever $ \langle h_i^{\dag}h_{i+1} \rangle$ and $\langle f_{i\sigma}^{\dag} f_{i+1\sigma}\rangle$ appear in the KS$^*$ formalism, Eq. (\ref{eq:hhff}) is automatically invoked. 

The values of $A$ and $B$ are determined as follows. %We will use $T_{KS^*}$ to denote the total kinetic energy of holons and spinons, and $t_{KS^*}$ as $T_{KS^*}$ divided by the total electron number. For a homogeneous chain ($V_i\equiv 0$, $n_i\equiv n$),
%\begin{eqnarray}
%    t_{KS^*}=-(2t^h_i\langle h_i^\dag h_{i+1}\rangle+4t^f_i\langle f_{i\sigma}^\dag f_{i+1\sigma}\rangle)/n 
%    \\ = -\frac{1}{n}((\frac{4t}{\pi}+4Bt(1-n)^2)\langle h_i^\dag h_{i+1}\rangle + (\frac{4J}{\pi}+ \nonumber \\
%    4At(1-n)+4BJ(1-n)^2)*\langle f_{i\sigma}^\dag f_{i+1\sigma}\rangle)
%\end{eqnarray}
%The factors ahead of $t_i^{h(f)}$ come from the complex conjugate and the spin degree of freedom. 
Different choices of $A$ and $B$ lead to different kinetic energies of the KS$^*$ auxiliary system ($t_{\mathrm{KS}^*}$), and, in turn, different XC energies ($\varepsilon_{\mathrm{xc}^*}$, as defined by the deference between $\varepsilon_{\mathrm{GS}}$ and $t_{\mathrm{KS}^*}$). In this sense, $A$ and $B$ tune the partition of the GS energy into the kinetic and potential parts, which is nonunique. In Fig.\ S3~\cite{SM}, we plot $t_{\mathrm{KS}^*}$ and $\varepsilon_{\mathrm{xc}^*}$ obtained from several values of $A$ and $B$, as well as from the rigorous mean-field definition of $\langle h_i^{\dag}h_{i+1} \rangle$ and $\langle f_{i\sigma}^{\dag} f_{i+1\sigma}\rangle$. Note that for homogeneous chains, they all consistently reproduce the exact GS energy. Testing calculations on inhomogeneous chains show that $A=0.45$ and $B=-0.05$ generally perform well. A phenomenological understanding is that this choice makes $t_{\mathrm{KS}^*}$ capture a predominant part of $\varepsilon_{\mathrm{GS}}$, and $\varepsilon_{\mathrm{xc}^*}$ shows a simple linear dependence on $n$ [Figs.\ \ref{fig:1}(a) and (b)]. In all the calculations below we will take the linear approximation of $\epsilon_{\mathrm{xc}^*}$ with fixed fitting parameters (see the caption of Fig. 1) without further fine-tuning.

$\hat H_{\mathrm{KS}^*}$ becomes formally exact by rederiving it as a variational minimum of the GS energy, i.e.~$\delta E_{\mathrm{GS}}/\delta \langle \Psi_{\mathrm{KS}^*}|=0$,
where $\langle \Psi_{\mathrm{KS}^*}|$ is the complex conjugate of the GS wave function of the KS$^*$ auxiliary system. %The form of  $V_{xc,i}$ appearing in Eq. (\ref{HKSS}) thus can be determined rigorously.

Special attention should be paid to the kinetic energy,  since $t_i^{f(h)}$ is now density dependent. We have
\begin{eqnarray}
\frac{\delta T_{\mathrm{KS}^*}}{\delta \langle \Psi_{\mathrm{KS}^*}|}&=&\hat T_{\mathrm{KS}^*}|\Psi_{\mathrm{KS}^*}\rangle \nonumber \\
&-&\langle \Psi_{\mathrm{KS}^*}|\frac{\partial \hat T_{\mathrm{KS}^*}}{\partial n_i}|\Psi_{\mathrm{KS}^*}\rangle \hat n_i |\Psi_{\mathrm{KS}^*}\rangle.
\end{eqnarray}
The second term on the right-hand side contributes a kinetically generated XC potential, which we denote as $V^T_{\mathrm{xc},i}\equiv\langle \Psi_{\mathrm{KS}^*}|\frac{\partial \hat T_{\mathrm{KS}^*}}{\partial n_i}|\Psi_{\mathrm{KS}^*}\rangle$. Under LDA, an explicit form can be given,
\begin{eqnarray}
    V^T_{\mathrm{xc},i}\approx&-&2\langle h_i^{\dag}h_{i+1} \rangle_{\mathrm{LDA}}\frac{d t_i^h}{dn}\bigg|_{n=n_i, \mathrm{LDA}} \nonumber \\
    &-&4\langle f_{i\sigma}^\dag f_{i+1\sigma}\rangle_{\mathrm{LDA}}\frac{d t_i^f}{dn}\bigg|_{n=n_i, \mathrm{LDA}}.  
\end{eqnarray}

Here, the subscript LDA means that to calculate the derivative at $n_i$, the LDA approximations of $\langle h_{i+1}^{\dag}h_{i}\rangle$ and $\langle f_{i+1\sigma}^{\dag} f_{i\sigma}\rangle$ [Eq.\ (11)] are invoked. In addition, $\delta E_{\mathrm{xc}^*}/\delta \langle\Psi_{\mathrm{KS}^*}|$ leads to the common interaction generated XC potential, which we denote as $V^I_{\mathrm{xc},i}$. The LDA form is
\begin{eqnarray}
    V^I_{\mathrm{xc},i}\approx\epsilon_{\mathrm{xc}^*}(n_i)+n_i\frac{d\epsilon_{\mathrm{xc}^*}}{dn}\bigg|_{n=n_i}.
\end{eqnarray}
The complete XC potential entering Eq.\ (\ref{HKSS}) is $V_{\mathrm{xc}}=V^T_{\mathrm{xc}}+V^I_{\mathrm{xc}}$.

We plot all the density functions in Fig.\ \ref{fig:1}(c). Plugging these into $\hat H_{\mathrm{KS}^*}$, self-consistent calculations can be performed to solve the $t$-$J$ chain with arbitrary input of lattice potentials ($V_{i=1,...,N_s})$ and number of doped holes ($N_h$). Figure \ref{fig:2} shows the computational flow chart, and Sec.\ IV of Supplemental Material~\cite{SM} contains more implementation details.

\begin{figure}
    \centering
    \includegraphics[width=0.45\textwidth]{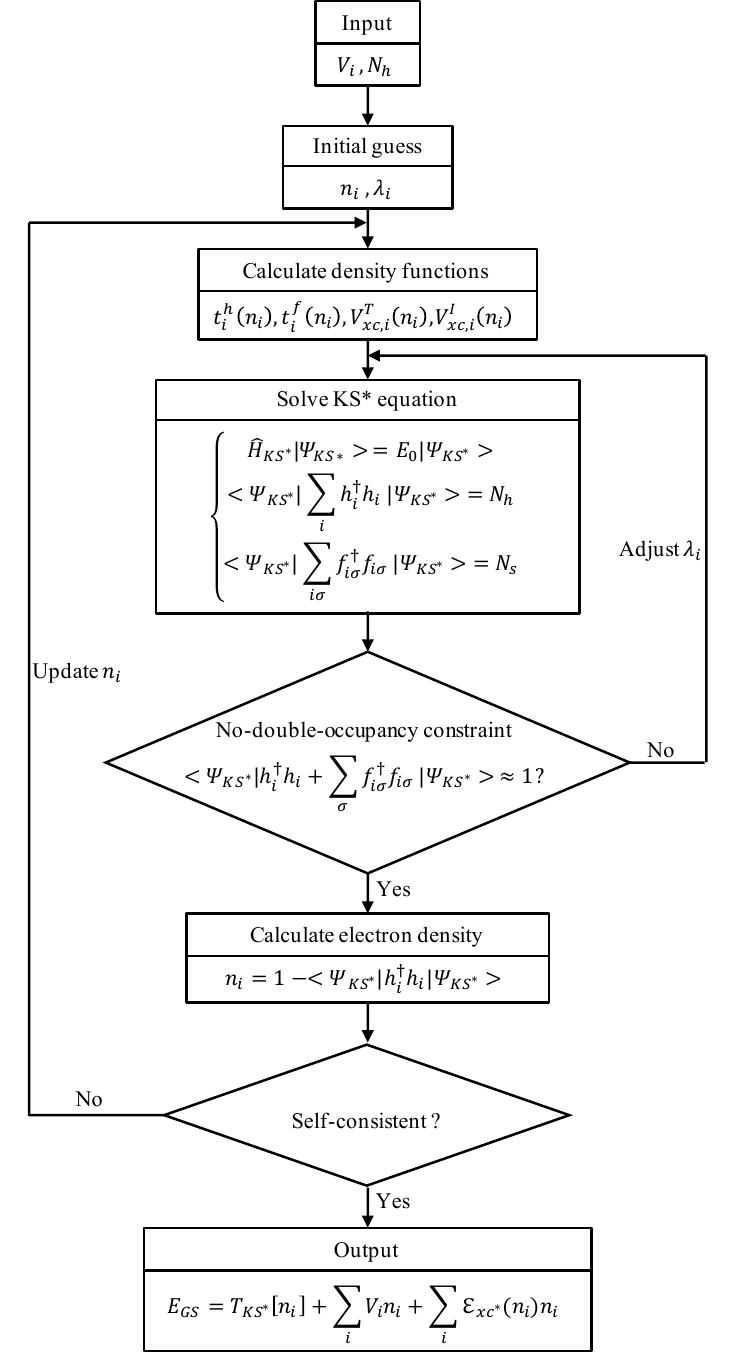}
    \caption{Schematic representation of the self-consistent iteration for solution of $\hat H_{\mathrm{KS}^*}$.}
    \label{fig:2}
\end{figure}

\begin{figure}[ht]
    \centering   
    \includegraphics[width=0.45\textwidth]{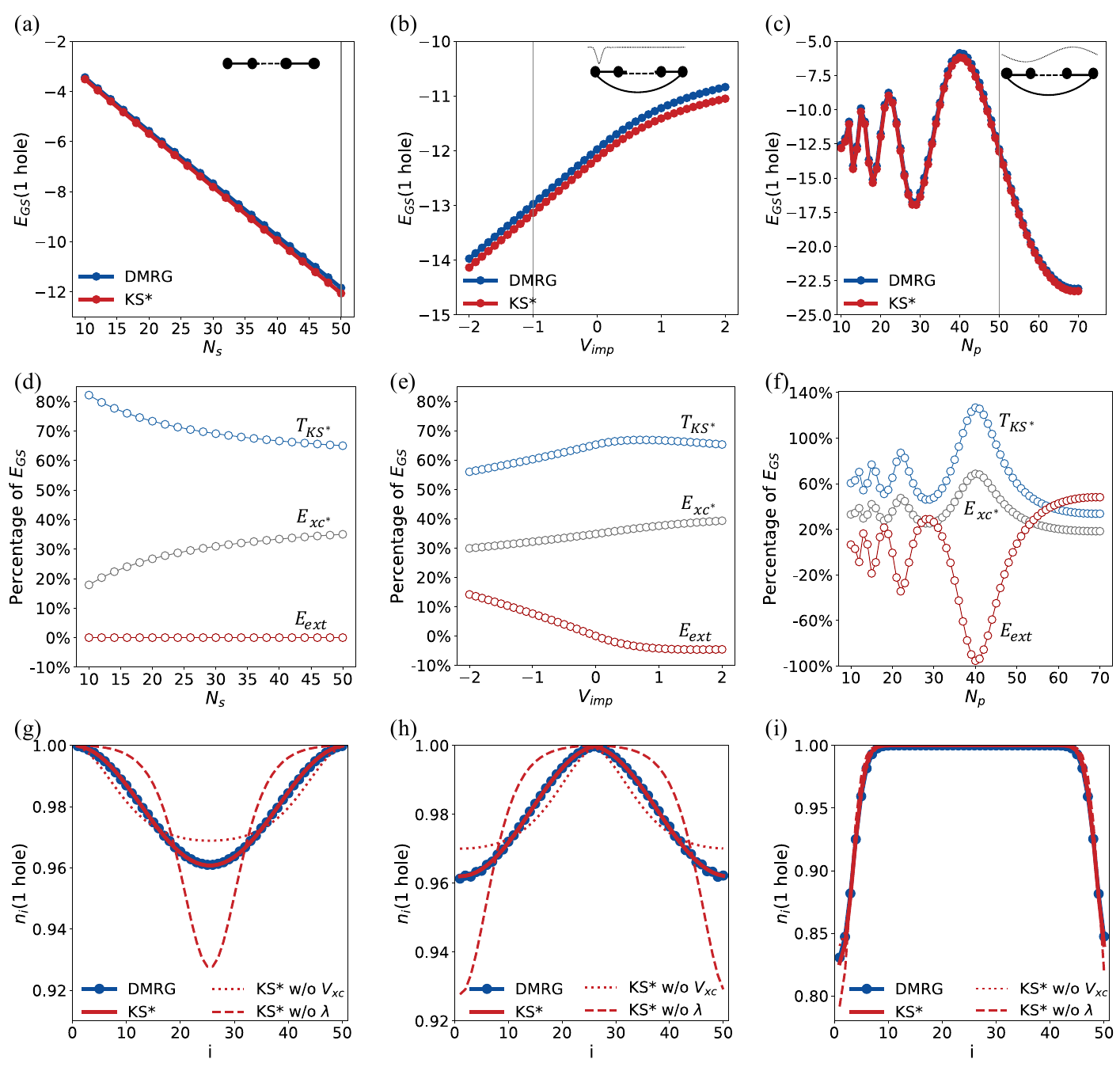}
    \caption{The one-hole GS energy of (a) a $N_s$-site OBC chain, (b) a 50-site PBC chain subject to a single-site impurity potential $V_i=\delta_{i1}V_{\mathrm{imp}}$, and (c) a 50-site PBC chain subject to a periodic potential $V_i =\cos(2\pi i/N_p)$. The vertical gray lines mark the parameter for the density plot in the bottom panels. (d)-(f) Decomposition of the calculated $E_{\mathrm{GS}}$ in the top panels into $T_{\mathrm{KS}^*}$, $E_{\mathrm{xc}^*}$, and $E_{\mathrm{ext}}\equiv \sum_i V_in_i$ within the KS$^*$ scheme. (g)-(i) The one-hole GS density distributions corresponding to the setups of the top panels. For (e), the impurity site is plotted as the center of the chain.}
    \label{fig:3}
\end{figure}

\begin{figure}[ht]
    \centering \includegraphics[width=0.43\textwidth]{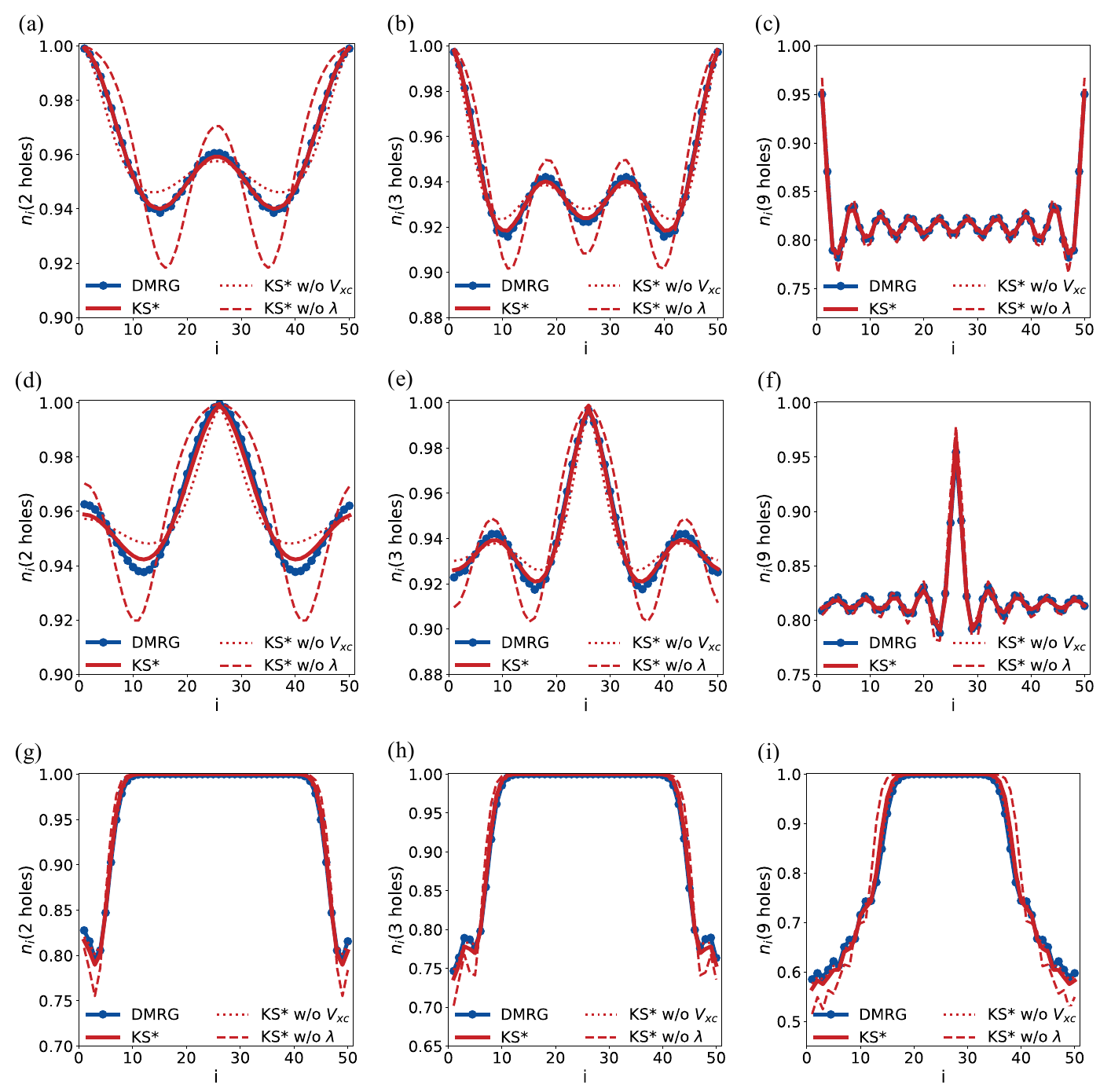}
    \caption{The two-, three- and nine-hole GS density distributions of (a)-(c) a 50-site OBC chain, (d)-(f) a 50-site PBC chain subject to a single-site impurity potential, and (g)-(i) a 50-site PBC chain subject to a periodic potential $V_{i}=\cos (2\pi i/N_p)$.}
    \label{fig:4}
\end{figure}

Figures~\ref{fig:3}(a)-\ref{fig:3}(c) summarize the one-hole GS energies in three representative cases with inhomogeneous density: (a) an OBC chain, (b) a PBC chain subject to a single-site impurity potential, and (c) a PBC chain subject to a periodic potential. $E_{\mathrm{GS}}$ is plotted as a function of (a) chain length ($N_s$), (b) impurity strength ($V_{\mathrm{imp}}$) and (c) potential period ($N_p$), respectively. For all the cases, we find that the KS$^*$ formalism is able to well trace the DMRG GS energy. More importantly, Figs.\ \ref{fig:3}(d)-\ref{fig:3}(f) decompose the calculated $E_{\mathrm{GS}}$  into $T_{\mathrm{KS}^*}$, $E_{\mathrm{xc}^*}$ and $E_{\mathrm{ext}}\equiv \sum_i V_in_i$ within the KS$^*$ scheme. Clearly, the total energy contains a non-negligible contribution from the XC energy and is crucial to obtain an accurate $E_{\mathrm{GS}}$, which, however, is missing in the slave-particle mean-field framework.

For each case, we select a parameter point and plot the GS density distribution in Figs.\ \ref{fig:3}(g)-\ref{fig:3}(i). We have also calculated the two-, three-, and nine-hole GS density distributions in Fig.\ \ref{fig:4}. All results show excellent agreement with the DMRG. While the local chemical potential $\lambda_i$ does not directly contribute to the energy, it is found to play an important role in improving the density description. 

The multihole energy comparison and decomposition are shown in Sec.\ V of Supplemental Material~\cite{SM}. It is worth mentioning that the DMRG algorithm implemented in ITensors~\cite{ITensor,ITensor-r0.3} takes more than 1 h to solve a 50-site $t$-$J$ chain using a 24-core high-performance computing node, while the KS$^*$ iteration finishes typically within 1 min on a laptop without parallel optimization.

In summary, we show that a DFT iteration algorithm formulated using the representation of emergent particles solves the 1D $t$-$J$ model with accuracy comparable to DMRG and much lower computational cost. The choice of an unconventional KS auxiliary system represents the key to extend the power of DFT to solve strongly correlated multiparticle quantum systems. For general spatial dimensions, one additional complexity is the emergent gauge field associated with fractionalization~\cite{Phasestring1,Phasestring2,Phasestring3,PRX22Weng}. We note that a recent closely related achievement is the demonstration of composite-fermion DFT for fractional quantum Hall systems~\cite{FQHDFT,FQHKS,ANYONKS}, which incorporates emergent gauge interactions in the KS auxiliary system by using the density-dependent flux attachment technique. Combining this technique with the KS$^*$ scheme will be a novel step to solve higher-dimensional strongly correlated models, and one step further, when we seek for DFT descriptions of realistic strongly correlated materials, such as high-$T_c$ cuprates. Another complexity is the competition of various distinct phases. A feasible strategy is similar to determining the mean-field phase diagram of a strongly correlated system. Suppose we now have a family of KS$^*$ equations, each reproducing the GS energy of a specific phase accurately. The lowest-energy solution thus selects the stable phase. On the whole, new versions of the KS scheme covering both fractionalized particles and emergent gauge fields are expected to carry DFT into a new realm.

\textit{Acknowledgments}—We would like to thank Z. Y. Weng for valuable discussions. This work is supported by National Natural Science Foundation of China (Grant No. 12374062, No. 92365201, No. 92477106, and No. 12074375), National Key R\&D Program of China (2023YFA1406400), Innovation Program for Quantum Science and Technology (2021ZD0301800) and the Fundamental Research Funds for the Central Universities.

%\bibliographystyle{unsrt}
%\bibliography{ref}

\begin{thebibliography}{45}%
\makeatletter
\providecommand \@ifxundefined [1]{%
 \@ifx{#1\undefined}
}%
\providecommand \@ifnum [1]{%
 \ifnum #1\expandafter \@firstoftwo
 \else \expandafter \@secondoftwo
 \fi
}%
\providecommand \@ifx [1]{%
 \ifx #1\expandafter \@firstoftwo
 \else \expandafter \@secondoftwo
 \fi
}%
\providecommand \natexlab [1]{#1}%
\providecommand \enquote  [1]{``#1''}%
\providecommand \bibnamefont  [1]{#1}%
\providecommand \bibfnamefont [1]{#1}%
\providecommand \citenamefont [1]{#1}%
\providecommand \href@noop [0]{\@secondoftwo}%
\providecommand \href [0]{\begingroup \@sanitize@url \@href}%
\providecommand \@href[1]{\@@startlink{#1}\@@href}%
\providecommand \@@href[1]{\endgroup#1\@@endlink}%
\providecommand \@sanitize@url [0]{\catcode `\\12\catcode `\$12\catcode `\&12\catcode `\#12\catcode `\^12\catcode `\_12\catcode `\%12\relax}%
\providecommand \@@startlink[1]{}%
\providecommand \@@endlink[0]{}%
\providecommand \url  [0]{\begingroup\@sanitize@url \@url }%
\providecommand \@url [1]{\endgroup\@href {#1}{\urlprefix }}%
\providecommand \urlprefix  [0]{URL }%
\providecommand \Eprint [0]{\href }%
\providecommand \doibase [0]{https://doi.org/}%
\providecommand \selectlanguage [0]{\@gobble}%
\providecommand \bibinfo  [0]{\@secondoftwo}%
\providecommand \bibfield  [0]{\@secondoftwo}%
\providecommand \translation [1]{[#1]}%
\providecommand \BibitemOpen [0]{}%
\providecommand \bibitemStop [0]{}%
\providecommand \bibitemNoStop [0]{.\EOS\space}%
\providecommand \EOS [0]{\spacefactor3000\relax}%
\providecommand \BibitemShut  [1]{\csname bibitem#1\endcsname}%
\let\auto@bib@innerbib\@empty
%</preamble>
\bibitem [{\citenamefont {Hohenberg}\ and\ \citenamefont {Kohn}(1964)}]{HK}%
  \BibitemOpen
  \bibfield  {author} {\bibinfo {author} {\bibfnamefont {P.}~\bibnamefont {Hohenberg}}\ and\ \bibinfo {author} {\bibfnamefont {W.}~\bibnamefont {Kohn}},\ }\bibfield  {title} {\bibinfo {title} {Inhomogeneous electron gas},\ }\href {https://doi.org/10.1103/PhysRev.136.B864} {\bibfield  {journal} {\bibinfo  {journal} {Phy. Rev. B}\ }\textbf {\bibinfo {volume} {136}},\ \bibinfo {pages} {B864} (\bibinfo {year} {1964})}\BibitemShut {NoStop}%
\bibitem [{\citenamefont {Jones}(2015)}]{DFT-review}%
  \BibitemOpen
  \bibfield  {author} {\bibinfo {author} {\bibfnamefont {R.~O.}\ \bibnamefont {Jones}},\ }\bibfield  {title} {\bibinfo {title} {Density functional theory: Its origins, rise to prominence, and future},\ }\href {https://doi.org/10.1103/RevModPhys.87.897} {\bibfield  {journal} {\bibinfo  {journal} {Rev. Mod. Phys.}\ }\textbf {\bibinfo {volume} {87}},\ \bibinfo {pages} {897} (\bibinfo {year} {2015})}\BibitemShut {NoStop}%
\bibitem [{\citenamefont {Kohn}\ and\ \citenamefont {Sham}(1965)}]{KS}%
  \BibitemOpen
  \bibfield  {author} {\bibinfo {author} {\bibfnamefont {W.}~\bibnamefont {Kohn}}\ and\ \bibinfo {author} {\bibfnamefont {L.~J.}\ \bibnamefont {Sham}},\ }\bibfield  {title} {\bibinfo {title} {Self-consistent equations including exchange and correlation effects},\ }\href {https://doi.org/10.1103/PhysRev.140.A1133} {\bibfield  {journal} {\bibinfo  {journal} {Phys. Rev.}\ }\textbf {\bibinfo {volume} {140}},\ \bibinfo {pages} {A1133} (\bibinfo {year} {1965})}\BibitemShut {NoStop}%
\bibitem [{\citenamefont {Kohn}(1999)}]{KohnReview}%
  \BibitemOpen
  \bibfield  {author} {\bibinfo {author} {\bibfnamefont {W.}~\bibnamefont {Kohn}},\ }\bibfield  {title} {\bibinfo {title} {Nobel lecture: Electronic structure of matter---wave functions and density functionals},\ }\href {https://doi.org/10.1103/RevModPhys.71.1253} {\bibfield  {journal} {\bibinfo  {journal} {Rev. Mod. Phys.}\ }\textbf {\bibinfo {volume} {71}},\ \bibinfo {pages} {1253} (\bibinfo {year} {1999})}\BibitemShut {NoStop}%
\bibitem [{\citenamefont {Gori-Giorgi}\ \emph {et~al.}(2009)\citenamefont {Gori-Giorgi}, \citenamefont {Seidl},\ and\ \citenamefont {Vignale}}]{strong-correlation-DFT-1}%
  \BibitemOpen
  \bibfield  {author} {\bibinfo {author} {\bibfnamefont {P.}~\bibnamefont {Gori-Giorgi}}, \bibinfo {author} {\bibfnamefont {M.}~\bibnamefont {Seidl}},\ and\ \bibinfo {author} {\bibfnamefont {G.}~\bibnamefont {Vignale}},\ }\bibfield  {title} {\bibinfo {title} {Density-functional theory for strongly interacting electrons},\ }\href {https://doi.org/10.1103/PhysRevLett.103.166402} {\bibfield  {journal} {\bibinfo  {journal} {Phys. Rev. Lett.}\ }\textbf {\bibinfo {volume} {103}},\ \bibinfo {pages} {166402} (\bibinfo {year} {2009})}\BibitemShut {NoStop}%
\bibitem [{\citenamefont {Malet}\ and\ \citenamefont {Gori-Giorgi}(2012)}]{strong-correlation-DFT-2}%
  \BibitemOpen
  \bibfield  {author} {\bibinfo {author} {\bibfnamefont {F.}~\bibnamefont {Malet}}\ and\ \bibinfo {author} {\bibfnamefont {P.}~\bibnamefont {Gori-Giorgi}},\ }\bibfield  {title} {\bibinfo {title} {{Strong correlation in Kohn-Sham density functional theory}},\ }\href {https://doi.org/10.1103/PhysRevLett.109.246402} {\bibfield  {journal} {\bibinfo  {journal} {Phys. Rev. Lett.}\ }\textbf {\bibinfo {volume} {109}},\ \bibinfo {pages} {246402} (\bibinfo {year} {2012})}\BibitemShut {NoStop}%
\bibitem [{\citenamefont {Kummel}\ and\ \citenamefont {Kronik}(2008)}]{ODDFT}%
  \BibitemOpen
  \bibfield  {author} {\bibinfo {author} {\bibfnamefont {S.}~\bibnamefont {Kummel}}\ and\ \bibinfo {author} {\bibfnamefont {L.}~\bibnamefont {Kronik}},\ }\bibfield  {title} {\bibinfo {title} {Orbital-dependent density functionals: Theory and applications},\ }\href {https://doi.org/10.1103/RevModPhys.80.3} {\bibfield  {journal} {\bibinfo  {journal} {Rev. Mod. Phys.}\ }\textbf {\bibinfo {volume} {80}},\ \bibinfo {pages} {3} (\bibinfo {year} {2008})}\BibitemShut {NoStop}%
\bibitem [{\citenamefont {Perdew}\ and\ \citenamefont {Schmidt}(2001)}]{perdew2001jacob}%
  \BibitemOpen
  \bibfield  {author} {\bibinfo {author} {\bibfnamefont {J.~P.}\ \bibnamefont {Perdew}}\ and\ \bibinfo {author} {\bibfnamefont {K.}~\bibnamefont {Schmidt}},\ }\bibfield  {title} {\bibinfo {title} {Jacob’s ladder of density functional approximations for the exchange-correlation energy},\ }\href {https://doi.org/https://doi.org/10.1063/1.1390175} {\bibfield  {journal} {\bibinfo  {journal} {AIP Conf. Proc.}\ }\textbf {\bibinfo {volume} {577}},\ \bibinfo {pages} {1} (\bibinfo {year} {2001})}\BibitemShut {NoStop}%
\bibitem [{\citenamefont {Sun}\ \emph {et~al.}(2015)\citenamefont {Sun}, \citenamefont {Ruzsinszky},\ and\ \citenamefont {Perdew}}]{SCAN}%
  \BibitemOpen
  \bibfield  {author} {\bibinfo {author} {\bibfnamefont {J.}~\bibnamefont {Sun}}, \bibinfo {author} {\bibfnamefont {A.}~\bibnamefont {Ruzsinszky}},\ and\ \bibinfo {author} {\bibfnamefont {J.~P.}\ \bibnamefont {Perdew}},\ }\bibfield  {title} {\bibinfo {title} {Strongly constrained and appropriately normed semilocal density functional},\ }\href {https://doi.org/10.1103/PhysRevLett.115.036402} {\bibfield  {journal} {\bibinfo  {journal} {Phys. Rev. Lett.}\ }\textbf {\bibinfo {volume} {115}},\ \bibinfo {pages} {036402} (\bibinfo {year} {2015})}\BibitemShut {NoStop}%
\bibitem [{\citenamefont {Kirkpatrick}\ \emph {et~al.}(2021)\citenamefont {Kirkpatrick}, \citenamefont {McMorrow}, \citenamefont {Turban}, \citenamefont {Gaunt}, \citenamefont {Spencer}, \citenamefont {Matthews}, \citenamefont {Obika}, \citenamefont {Thiry}, \citenamefont {Fortunato}, \citenamefont {Pfau}, \citenamefont {Castellanos}, \citenamefont {Petersen}, \citenamefont {Nelson}, \citenamefont {Kohli}, \citenamefont {Mori-Sánchez}, \citenamefont {Hassabis},\ and\ \citenamefont {Cohen}}]{DeepMind}%
  \BibitemOpen
  \bibfield  {author} {\bibinfo {author} {\bibfnamefont {J.}~\bibnamefont {Kirkpatrick}}, \bibinfo {author} {\bibfnamefont {B.}~\bibnamefont {McMorrow}}, \bibinfo {author} {\bibfnamefont {D.~H.~P.}\ \bibnamefont {Turban}}, \bibinfo {author} {\bibfnamefont {A.~L.}\ \bibnamefont {Gaunt}}, \bibinfo {author} {\bibfnamefont {J.~S.}\ \bibnamefont {Spencer}}, \bibinfo {author} {\bibfnamefont {A.~G. D.~G.}\ \bibnamefont {Matthews}}, \bibinfo {author} {\bibfnamefont {A.}~\bibnamefont {Obika}}, \bibinfo {author} {\bibfnamefont {L.}~\bibnamefont {Thiry}}, \bibinfo {author} {\bibfnamefont {M.}~\bibnamefont {Fortunato}}, \bibinfo {author} {\bibfnamefont {D.}~\bibnamefont {Pfau}}, \bibinfo {author} {\bibfnamefont {L.~R.}\ \bibnamefont {Castellanos}}, \bibinfo {author} {\bibfnamefont {S.}~\bibnamefont {Petersen}}, \bibinfo {author} {\bibfnamefont {A.~W.~R.}\ \bibnamefont {Nelson}}, \bibinfo {author} {\bibfnamefont {P.}~\bibnamefont {Kohli}}, \bibinfo {author} {\bibfnamefont {P.}~\bibnamefont {Mori-Sánchez}}, \bibinfo
  {author} {\bibfnamefont {D.}~\bibnamefont {Hassabis}},\ and\ \bibinfo {author} {\bibfnamefont {A.~J.}\ \bibnamefont {Cohen}},\ }\bibfield  {title} {\bibinfo {title} {Pushing the frontiers of density functionals by solving the fractional electron problem},\ }\href {https://doi.org/10.1126/science.abj6511} {\bibfield  {journal} {\bibinfo  {journal} {Science}\ }\textbf {\bibinfo {volume} {374}},\ \bibinfo {pages} {1385} (\bibinfo {year} {2021})}\BibitemShut {NoStop}%
\bibitem [{\citenamefont {Kotliar}\ \emph {et~al.}(2006)\citenamefont {Kotliar}, \citenamefont {Savrasov}, \citenamefont {Haule}, \citenamefont {Oudovenko}, \citenamefont {Parcollet},\ and\ \citenamefont {Marianetti}}]{DMFTReview}%
  \BibitemOpen
  \bibfield  {author} {\bibinfo {author} {\bibfnamefont {G.}~\bibnamefont {Kotliar}}, \bibinfo {author} {\bibfnamefont {S.~Y.}\ \bibnamefont {Savrasov}}, \bibinfo {author} {\bibfnamefont {K.}~\bibnamefont {Haule}}, \bibinfo {author} {\bibfnamefont {V.~S.}\ \bibnamefont {Oudovenko}}, \bibinfo {author} {\bibfnamefont {O.}~\bibnamefont {Parcollet}},\ and\ \bibinfo {author} {\bibfnamefont {C.~A.}\ \bibnamefont {Marianetti}},\ }\bibfield  {title} {\bibinfo {title} {Electronic structure calculations with dynamical mean-field theory},\ }\href {https://doi.org/10.1103/RevModPhys.78.865} {\bibfield  {journal} {\bibinfo  {journal} {Rev. Mod. Phys.}\ }\textbf {\bibinfo {volume} {78}},\ \bibinfo {pages} {865} (\bibinfo {year} {2006})}\BibitemShut {NoStop}%
\bibitem [{\citenamefont {Tomczak}\ \emph {et~al.}(2017)\citenamefont {Tomczak}, \citenamefont {Liu}, \citenamefont {Toschi}, \citenamefont {Kresse},\ and\ \citenamefont {Held}}]{GWDMFT}%
  \BibitemOpen
  \bibfield  {author} {\bibinfo {author} {\bibfnamefont {J.~M.}\ \bibnamefont {Tomczak}}, \bibinfo {author} {\bibfnamefont {P.}~\bibnamefont {Liu}}, \bibinfo {author} {\bibfnamefont {A.}~\bibnamefont {Toschi}}, \bibinfo {author} {\bibfnamefont {G.}~\bibnamefont {Kresse}},\ and\ \bibinfo {author} {\bibfnamefont {K.}~\bibnamefont {Held}},\ }\bibfield  {title} {\bibinfo {title} {{Merging GW with DMFT and non-local correlations beyond}},\ }\href {https://doi.org/10.1140/epjst/e2017-70053-1} {\bibfield  {journal} {\bibinfo  {journal} {Eur. Phys. J. Spec. Top.}\ }\textbf {\bibinfo {volume} {226}},\ \bibinfo {pages} {2565} (\bibinfo {year} {2017})}\BibitemShut {NoStop}%
\bibitem [{\citenamefont {Kang}\ \emph {et~al.}()\citenamefont {Kang}, \citenamefont {Semon}, \citenamefont {Melnick}, \citenamefont {Kotliar},\ and\ \citenamefont {Choi}}]{comdmftv2}%
  \BibitemOpen
  \bibfield  {author} {\bibinfo {author} {\bibfnamefont {B.}~\bibnamefont {Kang}}, \bibinfo {author} {\bibfnamefont {P.}~\bibnamefont {Semon}}, \bibinfo {author} {\bibfnamefont {C.}~\bibnamefont {Melnick}}, \bibinfo {author} {\bibfnamefont {G.}~\bibnamefont {Kotliar}},\ and\ \bibinfo {author} {\bibfnamefont {S.}~\bibnamefont {Choi}},\ }\bibfield  {title} {\bibinfo {title} {{ComDMFT v.2.0: Fully self-consistent ab initio GW+EDMFT for the electronic structure of correlated quantum materials}},\ }\href {https://arxiv.org/abs/2310.04613} {\bibinfo  {journal} {arXiv:2310.04613}\ }\BibitemShut {NoStop}%
\bibitem [{\citenamefont {Onida}\ \emph {et~al.}(2002)\citenamefont {Onida}, \citenamefont {Reining},\ and\ \citenamefont {Rubio}}]{GWreview}%
  \BibitemOpen
\bibfield  {journal} {  }\bibfield  {author} {\bibinfo {author} {\bibfnamefont {G.}~\bibnamefont {Onida}}, \bibinfo {author} {\bibfnamefont {L.}~\bibnamefont {Reining}},\ and\ \bibinfo {author} {\bibfnamefont {A.}~\bibnamefont {Rubio}},\ }\bibfield  {title} {\bibinfo {title} {{Electronic excitations: Density-functional versus many-body Green’s-function approaches}},\ }\href {https://doi.org/10.1103/RevModPhys.74.601} {\bibfield  {journal} {\bibinfo  {journal} {Rev. Mod. Phys.}\ }\textbf {\bibinfo {volume} {74}},\ \bibinfo {pages} {601} (\bibinfo {year} {2002})}\BibitemShut {NoStop}%
\bibitem [{\citenamefont {Sachdev}(2023)}]{SachdevQPM}%
  \BibitemOpen
  \bibfield  {author} {\bibinfo {author} {\bibfnamefont {S.}~\bibnamefont {Sachdev}},\ }\href {https://doi.org/10.1017/9781009212717} {\emph {\bibinfo {title} {Quantum Phases of Matter}}}\ (\bibinfo  {publisher} {Cambridge University Press, Cambridge, England},\ \bibinfo {year} {2023})\BibitemShut {NoStop}%
\bibitem [{\citenamefont {Voit}(2000)}]{LuttingerReview}%
  \BibitemOpen
  \bibfield  {author} {\bibinfo {author} {\bibfnamefont {J.}~\bibnamefont {Voit}},\ }\bibfield  {title} {\bibinfo {title} {{A brief introduction to Luttinger liquids}},\ }\href {https://doi.org/10.1063/1.1342524} {\bibfield  {journal} {\bibinfo  {journal} {AIP Conf. Proc.}\ }\textbf {\bibinfo {volume} {544}},\ \bibinfo {pages} {309} (\bibinfo {year} {2000})}\BibitemShut {NoStop}%
\bibitem [{\citenamefont {Lee}\ \emph {et~al.}(2006)\citenamefont {Lee}, \citenamefont {Nagaosa},\ and\ \citenamefont {Wen}}]{DopeMott}%
  \BibitemOpen
  \bibfield  {author} {\bibinfo {author} {\bibfnamefont {P.~A.}\ \bibnamefont {Lee}}, \bibinfo {author} {\bibfnamefont {N.}~\bibnamefont {Nagaosa}},\ and\ \bibinfo {author} {\bibfnamefont {X.-G.}\ \bibnamefont {Wen}},\ }\bibfield  {title} {\bibinfo {title} {{Doping a Mott insulator: Physics of high-temperature superconductivity}},\ }\href {https://doi.org/10.1103/RevModPhys.78.17} {\bibfield  {journal} {\bibinfo  {journal} {Rev. Mod. Phys.}\ }\textbf {\bibinfo {volume} {78}},\ \bibinfo {pages} {17} (\bibinfo {year} {2006})}\BibitemShut {NoStop}%
\bibitem [{\citenamefont {Capelle}\ and\ \citenamefont {Campo~Jr}(2013)}]{capelle2013density}%
  \BibitemOpen
  \bibfield  {author} {\bibinfo {author} {\bibfnamefont {K.}~\bibnamefont {Capelle}}\ and\ \bibinfo {author} {\bibfnamefont {V.~L.}\ \bibnamefont {Campo~Jr}},\ }\bibfield  {title} {\bibinfo {title} {{Density functionals and model Hamiltonians: Pillars of many-particle physics}},\ }\href {https://doi.org/https://doi.org/10.1016/j.physrep.2013.03.002} {\bibfield  {journal} {\bibinfo  {journal} {Phy. Rep.}\ }\textbf {\bibinfo {volume} {528}},\ \bibinfo {pages} {91} (\bibinfo {year} {2013})}\BibitemShut {NoStop}%
\bibitem [{\citenamefont {Lieb}\ and\ \citenamefont {Wu}(1968)}]{LiebWu}%
  \BibitemOpen
  \bibfield  {author} {\bibinfo {author} {\bibfnamefont {E.~H.}\ \bibnamefont {Lieb}}\ and\ \bibinfo {author} {\bibfnamefont {F.~Y.}\ \bibnamefont {Wu}},\ }\bibfield  {title} {\bibinfo {title} {{Absence of Mott transition in an exact solution of the short-range, one-band model in one dimension}},\ }\href {https://doi.org/10.1103/PhysRevLett.20.1445} {\bibfield  {journal} {\bibinfo  {journal} {Phys. Rev. Lett.}\ }\textbf {\bibinfo {volume} {20}},\ \bibinfo {pages} {1445} (\bibinfo {year} {1968})}\BibitemShut {NoStop}%
\bibitem [{\citenamefont {Schollw\"ock}(2005)}]{DMRG}%
  \BibitemOpen
  \bibfield  {author} {\bibinfo {author} {\bibfnamefont {U.}~\bibnamefont {Schollw\"ock}},\ }\bibfield  {title} {\bibinfo {title} {The density-matrix renormalization group},\ }\href {https://doi.org/10.1103/RevModPhys.77.259} {\bibfield  {journal} {\bibinfo  {journal} {Rev. Mod. Phys.}\ }\textbf {\bibinfo {volume} {77}},\ \bibinfo {pages} {259} (\bibinfo {year} {2005})}\BibitemShut {NoStop}%
\bibitem [{\citenamefont {L\'evy}(1979)}]{levy1979}%
  \BibitemOpen
  \bibfield  {author} {\bibinfo {author} {\bibfnamefont {M.}~\bibnamefont {L\'evy}},\ }\bibfield  {title} {\bibinfo {title} {Universal variational functionals of electron densities, first-order density matrices, and natural spin-orbitals and solution of the v-representability problem},\ }\href {https://doi.org/https://doi.org/10.1073/pnas.76.12.6062} {\bibfield  {journal} {\bibinfo  {journal} {Proc. Natl. Acad. Sci. U.S.A.}\ }\textbf {\bibinfo {volume} {76}},\ \bibinfo {pages} {6062} (\bibinfo {year} {1979})}\BibitemShut {NoStop}%
\bibitem [{\citenamefont {Lieb}(1983)}]{lieb1983}%
  \BibitemOpen
  \bibfield  {author} {\bibinfo {author} {\bibfnamefont {E.~H.}\ \bibnamefont {Lieb}},\ }\bibfield  {title} {\bibinfo {title} {{Density functionals for Coulomb systems}},\ }\href {https://doi.org/https://doi.org/10.1002/qua.560240302} {\bibfield  {journal} {\bibinfo  {journal} {Int. J. Quantum Chem.}\ }\textbf {\bibinfo {volume} {24}},\ \bibinfo {pages} {243} (\bibinfo {year} {1983})}\BibitemShut {NoStop}%
\bibitem [{\citenamefont {Chayes}\ \emph {et~al.}(1985)\citenamefont {Chayes}, \citenamefont {Chayes},\ and\ \citenamefont {Ruskai}}]{chayes1985density}%
  \BibitemOpen
  \bibfield  {author} {\bibinfo {author} {\bibfnamefont {J.~T.}\ \bibnamefont {Chayes}}, \bibinfo {author} {\bibfnamefont {L.}~\bibnamefont {Chayes}},\ and\ \bibinfo {author} {\bibfnamefont {M.~B.}\ \bibnamefont {Ruskai}},\ }\bibfield  {title} {\bibinfo {title} {Density functional approach to quantum lattice systems},\ }\href {https://doi.org/https://doi.org/10.1007/BF01010474} {\bibfield  {journal} {\bibinfo  {journal} {J. Stat. Phys.}\ }\textbf {\bibinfo {volume} {38}},\ \bibinfo {pages} {497} (\bibinfo {year} {1985})}\BibitemShut {NoStop}%
\bibitem [{\citenamefont {Penz}\ and\ \citenamefont {van Leeuwen}(2024)}]{penz2024geometrical}%
  \BibitemOpen
  \bibfield  {author} {\bibinfo {author} {\bibfnamefont {M.}~\bibnamefont {Penz}}\ and\ \bibinfo {author} {\bibfnamefont {R.}~\bibnamefont {van Leeuwen}},\ }\bibfield  {title} {\bibinfo {title} {Geometrical perspective on spin–lattice density-functional theory},\ }\href {https://doi.org/10.1063/5.0230494} {\bibfield  {journal} {\bibinfo  {journal} {J. Chem. Phys.}\ }\textbf {\bibinfo {volume} {161}},\ \bibinfo {pages} {150901} (\bibinfo {year} {2024})}\BibitemShut {NoStop}%
\bibitem [{\citenamefont {Mao}\ \emph {et~al.}(2021)\citenamefont {Mao}, \citenamefont {Tang}, \citenamefont {Duan},\ and\ \citenamefont {Liu}}]{maoPRB}%
  \BibitemOpen
  \bibfield  {author} {\bibinfo {author} {\bibfnamefont {J.}~\bibnamefont {Mao}}, \bibinfo {author} {\bibfnamefont {H.}~\bibnamefont {Tang}}, \bibinfo {author} {\bibfnamefont {W.}~\bibnamefont {Duan}},\ and\ \bibinfo {author} {\bibfnamefont {Z.}~\bibnamefont {Liu}},\ }\bibfield  {title} {\bibinfo {title} {{Testing density functional theory in a quantum Ising chain}},\ }\href {https://doi.org/10.1103/PhysRevB.104.155145} {\bibfield  {journal} {\bibinfo  {journal} {Phys. Rev. B}\ }\textbf {\bibinfo {volume} {104}},\ \bibinfo {pages} {155145} (\bibinfo {year} {2021})}\BibitemShut {NoStop}%
\bibitem [{\citenamefont {Xu}\ \emph {et~al.}(2022)\citenamefont {Xu}, \citenamefont {Mao}, \citenamefont {Gao},\ and\ \citenamefont {Liu}}]{xuliming}%
  \BibitemOpen
  \bibfield  {author} {\bibinfo {author} {\bibfnamefont {L.}~\bibnamefont {Xu}}, \bibinfo {author} {\bibfnamefont {J.}~\bibnamefont {Mao}}, \bibinfo {author} {\bibfnamefont {X.}~\bibnamefont {Gao}},\ and\ \bibinfo {author} {\bibfnamefont {Z.}~\bibnamefont {Liu}},\ }\bibfield  {title} {\bibinfo {title} {{Extensibility of Hohenberg–Kohn theorem to general quantum systems}},\ }\href {https://doi.org/https://doi.org/10.1002/qute.202200041} {\bibfield  {journal} {\bibinfo  {journal} {Adv. Quantum Technol.}\ }\textbf {\bibinfo {volume} {5}},\ \bibinfo {pages} {2200041} (\bibinfo {year} {2022})}\BibitemShut {NoStop}%
\bibitem [{\citenamefont {Ceperley}\ and\ \citenamefont {Alder}(1980)}]{LDACA}%
  \BibitemOpen
  \bibfield  {author} {\bibinfo {author} {\bibfnamefont {D.~M.}\ \bibnamefont {Ceperley}}\ and\ \bibinfo {author} {\bibfnamefont {B.~J.}\ \bibnamefont {Alder}},\ }\bibfield  {title} {\bibinfo {title} {Ground state of the electron gas by a stochastic method},\ }\href {https://doi.org/10.1103/PhysRevLett.45.566} {\bibfield  {journal} {\bibinfo  {journal} {Phys. Rev. Lett.}\ }\textbf {\bibinfo {volume} {45}},\ \bibinfo {pages} {566} (\bibinfo {year} {1980})}\BibitemShut {NoStop}%
\bibitem [{\citenamefont {White}(1993)}]{DMRG1993}%
  \BibitemOpen
  \bibfield  {author} {\bibinfo {author} {\bibfnamefont {S.~R.}\ \bibnamefont {White}},\ }\bibfield  {title} {\bibinfo {title} {Density-matrix algorithms for quantum renormalization groups},\ }\href {https://doi.org/10.1103/PhysRevB.48.10345} {\bibfield  {journal} {\bibinfo  {journal} {Phys. Rev. B}\ }\textbf {\bibinfo {volume} {48}},\ \bibinfo {pages} {10345} (\bibinfo {year} {1993})}\BibitemShut {NoStop}%
\bibitem [{\citenamefont {Schollw{\"o}ck}(2011)}]{DMRG2011}%
  \BibitemOpen
  \bibfield  {author} {\bibinfo {author} {\bibfnamefont {U.}~\bibnamefont {Schollw{\"o}ck}},\ }\bibfield  {title} {\bibinfo {title} {The density-matrix renormalization group in the age of matrix product states},\ }\href {https://doi.org/10.1016/j.aop.2010.09.012} {\bibfield  {journal} {\bibinfo  {journal} {Ann. Phys. (Amsterdam)}\ }\textbf {\bibinfo {volume} {326}},\ \bibinfo {pages} {96} (\bibinfo {year} {2011})}\BibitemShut {NoStop}%
\bibitem [{\citenamefont {Moreno}\ \emph {et~al.}(2011)\citenamefont {Moreno}, \citenamefont {Muramatsu},\ and\ \citenamefont {Manmana}}]{phase-diagram-1DtJ}%
  \BibitemOpen
  \bibfield  {author} {\bibinfo {author} {\bibfnamefont {A.}~\bibnamefont {Moreno}}, \bibinfo {author} {\bibfnamefont {A.}~\bibnamefont {Muramatsu}},\ and\ \bibinfo {author} {\bibfnamefont {S.~R.}\ \bibnamefont {Manmana}},\ }\bibfield  {title} {\bibinfo {title} {{Ground-state phase diagram of the one-dimensional $t-J$ model}},\ }\href {https://doi.org/10.1103/PhysRevB.83.205113} {\bibfield  {journal} {\bibinfo  {journal} {Phys. Rev. B}\ }\textbf {\bibinfo {volume} {83}},\ \bibinfo {pages} {205113} (\bibinfo {year} {2011})}\BibitemShut {NoStop}%
\bibitem [{\citenamefont {Zhu}\ \emph {et~al.}(2016)\citenamefont {Zhu}, \citenamefont {Wang}, \citenamefont {Sheng},\ and\ \citenamefont {Weng}}]{NPB16Zhu}%
  \BibitemOpen
  \bibfield  {author} {\bibinfo {author} {\bibfnamefont {Z.}~\bibnamefont {Zhu}}, \bibinfo {author} {\bibfnamefont {Q.-R.}\ \bibnamefont {Wang}}, \bibinfo {author} {\bibfnamefont {D.}~\bibnamefont {Sheng}},\ and\ \bibinfo {author} {\bibfnamefont {Z.-Y.}\ \bibnamefont {Weng}},\ }\bibfield  {title} {\bibinfo {title} {{Exact sign structure of the $t-J$ chain and the single hole ground state}},\ }\href {https://doi.org/https://doi.org/10.1016/j.nuclphysb.2015.12.004} {\bibfield  {journal} {\bibinfo  {journal} {Nucl. Phys. B}\ }\textbf {\bibinfo {volume} {903}},\ \bibinfo {pages} {51} (\bibinfo {year} {2016})}\BibitemShut {NoStop}%
\bibitem [{\citenamefont {Coulthard}\ \emph {et~al.}(2018)\citenamefont {Coulthard}, \citenamefont {Clark},\ and\ \citenamefont {Jaksch}}]{phase-diagram-1DtJ-pairhoppind}%
  \BibitemOpen
  \bibfield  {author} {\bibinfo {author} {\bibfnamefont {J.~R.}\ \bibnamefont {Coulthard}}, \bibinfo {author} {\bibfnamefont {S.~R.}\ \bibnamefont {Clark}},\ and\ \bibinfo {author} {\bibfnamefont {D.}~\bibnamefont {Jaksch}},\ }\bibfield  {title} {\bibinfo {title} {{Ground-state phase diagram of the one-dimensional $t-J$ model with pair hopping terms}},\ }\href {https://doi.org/10.1103/PhysRevB.98.035116} {\bibfield  {journal} {\bibinfo  {journal} {Phys. Rev. B}\ }\textbf {\bibinfo {volume} {98}},\ \bibinfo {pages} {035116} (\bibinfo {year} {2018})}\BibitemShut {NoStop}%
\bibitem [{\citenamefont {Zhao}\ \emph {et~al.}(2022)\citenamefont {Zhao}, \citenamefont {Chen}, \citenamefont {Zhang},\ and\ \citenamefont {Weng}}]{PRX22Weng}%
  \BibitemOpen
  \bibfield  {author} {\bibinfo {author} {\bibfnamefont {J.-Y.}\ \bibnamefont {Zhao}}, \bibinfo {author} {\bibfnamefont {S.~A.}\ \bibnamefont {Chen}}, \bibinfo {author} {\bibfnamefont {H.-K.}\ \bibnamefont {Zhang}},\ and\ \bibinfo {author} {\bibfnamefont {Z.-Y.}\ \bibnamefont {Weng}},\ }\bibfield  {title} {\bibinfo {title} {Two-hole ground state: Dichotomy in pairing symmetry},\ }\href {https://doi.org/10.1103/PhysRevX.12.011062} {\bibfield  {journal} {\bibinfo  {journal} {Phys. Rev. X}\ }\textbf {\bibinfo {volume} {12}},\ \bibinfo {pages} {011062} (\bibinfo {year} {2022})}\BibitemShut {NoStop}%
\bibitem [{SM()}]{SM}%
  \BibitemOpen
  \href@noop {} {}\bibinfo {note} {See Supplemental Material at \href{https://doi.org/10.1103/PhysRevLett.134.136505}{http://link.aps.org/ supplemental/10.1103/PhysRevLett.134.136505} for implementation details, additional computational results and benchmark tests.}\BibitemShut {Stop}%
\bibitem [{\citenamefont {Deng}\ \emph {et~al.}(2009)\citenamefont {Deng}, \citenamefont {Wang}, \citenamefont {Dai},\ and\ \citenamefont {Fang}}]{PRB09Gutzwiller}%
  \BibitemOpen
  \bibfield  {author} {\bibinfo {author} {\bibfnamefont {X.~Y.}\ \bibnamefont {Deng}}, \bibinfo {author} {\bibfnamefont {L.}~\bibnamefont {Wang}}, \bibinfo {author} {\bibfnamefont {X.}~\bibnamefont {Dai}},\ and\ \bibinfo {author} {\bibfnamefont {Z.}~\bibnamefont {Fang}},\ }\bibfield  {title} {\bibinfo {title} {{Local density approximation combined with Gutzwiller method for correlated electron systems: Formalism and applications}},\ }\href {https://doi.org/10.1103/PhysRevB.79.075114} {\bibfield  {journal} {\bibinfo  {journal} {Phys. Rev. B}\ }\textbf {\bibinfo {volume} {79}},\ \bibinfo {pages} {075114} (\bibinfo {year} {2009})}\BibitemShut {NoStop}%
\bibitem [{\citenamefont {Ye}\ \emph {et~al.}(2022)\citenamefont {Ye}, \citenamefont {Fang}, \citenamefont {Zhang}, \citenamefont {Zhang}, \citenamefont {Wu}, \citenamefont {Lu}, \citenamefont {Yao}, \citenamefont {Wang},\ and\ \citenamefont {Ho}}]{JPC22Gutzwiller}%
  \BibitemOpen
  \bibfield  {author} {\bibinfo {author} {\bibfnamefont {Z.}~\bibnamefont {Ye}}, \bibinfo {author} {\bibfnamefont {Y.}~\bibnamefont {Fang}}, \bibinfo {author} {\bibfnamefont {H.}~\bibnamefont {Zhang}}, \bibinfo {author} {\bibfnamefont {F.}~\bibnamefont {Zhang}}, \bibinfo {author} {\bibfnamefont {S.}~\bibnamefont {Wu}}, \bibinfo {author} {\bibfnamefont {W.-C.}\ \bibnamefont {Lu}}, \bibinfo {author} {\bibfnamefont {Y.-X.}\ \bibnamefont {Yao}}, \bibinfo {author} {\bibfnamefont {C.-Z.}\ \bibnamefont {Wang}},\ and\ \bibinfo {author} {\bibfnamefont {K.-M.}\ \bibnamefont {Ho}},\ }\bibfield  {title} {\bibinfo {title} {{The Gutzwiller conjugate gradient minimization method for correlated electron systems}},\ }\href {https://doi.org/10.1088/1361-648X/ac5e03} {\bibfield  {journal} {\bibinfo  {journal} {J. Phys. C}\ }\textbf {\bibinfo {volume} {34}},\ \bibinfo {pages} {243001} (\bibinfo {year} {2022})}\BibitemShut {NoStop}%
\bibitem [{\citenamefont {Haldane}(1981)}]{HaldanLuttinger}%
  \BibitemOpen
  \bibfield  {author} {\bibinfo {author} {\bibfnamefont {F.~D.~M.}\ \bibnamefont {Haldane}},\ }\bibfield  {title} {\bibinfo {title} {{Luttinger liquid theory of one-dimensional quantum fluids. I. Properties of the Luttinger model and their extension to the general 1D interacting spinless Fermi gas}},\ }\href {https://doi.org/10.1088/0022-3719/14/19/010} {\bibfield  {journal} {\bibinfo  {journal} {J. Phys. C}\ }\textbf {\bibinfo {volume} {14}},\ \bibinfo {pages} {2585} (\bibinfo {year} {1981})}\BibitemShut {NoStop}%
\bibitem [{\citenamefont {Weng}\ \emph {et~al.}(1995)\citenamefont {Weng}, \citenamefont {Sheng},\ and\ \citenamefont {Ting}}]{Phasestring1}%
  \BibitemOpen
  \bibfield  {author} {\bibinfo {author} {\bibfnamefont {Z.~Y.}\ \bibnamefont {Weng}}, \bibinfo {author} {\bibfnamefont {D.~N.}\ \bibnamefont {Sheng}},\ and\ \bibinfo {author} {\bibfnamefont {C.~S.}\ \bibnamefont {Ting}},\ }\bibfield  {title} {\bibinfo {title} {{Spin-charge separation in the $t-J$ model: Magnetic and transport anomalies}},\ }\href {https://doi.org/10.1103/PhysRevB.52.637} {\bibfield  {journal} {\bibinfo  {journal} {Phys. Rev. B}\ }\textbf {\bibinfo {volume} {52}},\ \bibinfo {pages} {637} (\bibinfo {year} {1995})}\BibitemShut {NoStop}%
\bibitem [{\citenamefont {Weng}\ \emph {et~al.}(1997)\citenamefont {Weng}, \citenamefont {Sheng}, \citenamefont {Chen},\ and\ \citenamefont {Ting}}]{Phasestring2}%
  \BibitemOpen
  \bibfield  {author} {\bibinfo {author} {\bibfnamefont {Z.~Y.}\ \bibnamefont {Weng}}, \bibinfo {author} {\bibfnamefont {D.~N.}\ \bibnamefont {Sheng}}, \bibinfo {author} {\bibfnamefont {Y.-C.}\ \bibnamefont {Chen}},\ and\ \bibinfo {author} {\bibfnamefont {C.~S.}\ \bibnamefont {Ting}},\ }\bibfield  {title} {\bibinfo {title} {{Phase string effect in the $t-J$ model: General theory}},\ }\href {https://doi.org/10.1103/PhysRevB.55.3894} {\bibfield  {journal} {\bibinfo  {journal} {Phys. Rev. B}\ }\textbf {\bibinfo {volume} {55}},\ \bibinfo {pages} {3894} (\bibinfo {year} {1997})}\BibitemShut {NoStop}%
\bibitem [{\citenamefont {Fishman}\ \emph {et~al.}(2022{\natexlab{a}})\citenamefont {Fishman}, \citenamefont {White},\ and\ \citenamefont {Stoudenmire}}]{ITensor}%
  \BibitemOpen
  \bibfield  {author} {\bibinfo {author} {\bibfnamefont {M.}~\bibnamefont {Fishman}}, \bibinfo {author} {\bibfnamefont {S.~R.}\ \bibnamefont {White}},\ and\ \bibinfo {author} {\bibfnamefont {E.~M.}\ \bibnamefont {Stoudenmire}},\ }\bibfield  {title} {\bibinfo {title} {{The ITensor software library for tensor network calculations}},\ }\href {https://doi.org/10.21468/SciPostPhysCodeb.4} {\bibfield  {journal} {\bibinfo  {journal} {SciPost Phys. Codebases}\ ,\ \bibinfo {pages} {4}} (\bibinfo {year} {2022}{\natexlab{a}})}\BibitemShut {NoStop}%
\bibitem [{\citenamefont {Fishman}\ \emph {et~al.}(2022{\natexlab{b}})\citenamefont {Fishman}, \citenamefont {White},\ and\ \citenamefont {Stoudenmire}}]{ITensor-r0.3}%
  \BibitemOpen
  \bibfield  {author} {\bibinfo {author} {\bibfnamefont {M.}~\bibnamefont {Fishman}}, \bibinfo {author} {\bibfnamefont {S.~R.}\ \bibnamefont {White}},\ and\ \bibinfo {author} {\bibfnamefont {E.~M.}\ \bibnamefont {Stoudenmire}},\ }\bibfield  {title} {\bibinfo {title} {{Codebase release 0.3 for ITensor}},\ }\href {https://doi.org/10.21468/SciPostPhysCodeb.4-r0.3} {\bibfield  {journal} {\bibinfo  {journal} {SciPost Phys. Codebases}\ ,\ \bibinfo {pages} {4}} (\bibinfo {year} {2022}{\natexlab{b}})}\BibitemShut {NoStop}%
\bibitem [{\citenamefont {Zhao}\ and\ \citenamefont {Weng}(2022)}]{Phasestring3}%
  \BibitemOpen
  \bibfield  {author} {\bibinfo {author} {\bibfnamefont {J.-Y.}\ \bibnamefont {Zhao}}\ and\ \bibinfo {author} {\bibfnamefont {Z.-Y.}\ \bibnamefont {Weng}},\ }\bibfield  {title} {\bibinfo {title} {{Mottness, phase string, and high-Tc superconductivity}},\ }\href {https://doi.org/10.1088/1674-1056/ac7a14} {\bibfield  {journal} {\bibinfo  {journal} {Chin. Phys. B}\ }\textbf {\bibinfo {volume} {31}},\ \bibinfo {pages} {087104} (\bibinfo {year} {2022})}\BibitemShut {NoStop}%
\bibitem [{\citenamefont {Zhao}\ \emph {et~al.}(2017)\citenamefont {Zhao}, \citenamefont {Thakurathi}, \citenamefont {Jain}, \citenamefont {Sen},\ and\ \citenamefont {Jain}}]{FQHDFT}%
  \BibitemOpen
  \bibfield  {author} {\bibinfo {author} {\bibfnamefont {J.}~\bibnamefont {Zhao}}, \bibinfo {author} {\bibfnamefont {M.}~\bibnamefont {Thakurathi}}, \bibinfo {author} {\bibfnamefont {M.}~\bibnamefont {Jain}}, \bibinfo {author} {\bibfnamefont {D.}~\bibnamefont {Sen}},\ and\ \bibinfo {author} {\bibfnamefont {J.~K.}\ \bibnamefont {Jain}},\ }\bibfield  {title} {\bibinfo {title} {{Density-functional theory of the fractional quantum Hall effect}},\ }\href {https://doi.org/10.1103/PhysRevLett.118.196802} {\bibfield  {journal} {\bibinfo  {journal} {Phys. Rev. Lett.}\ }\textbf {\bibinfo {volume} {118}},\ \bibinfo {pages} {196802} (\bibinfo {year} {2017})}\BibitemShut {NoStop}%
\bibitem [{\citenamefont {Hu}\ and\ \citenamefont {Jain}(2019)}]{FQHKS}%
  \BibitemOpen
  \bibfield  {author} {\bibinfo {author} {\bibfnamefont {Y.}~\bibnamefont {Hu}}\ and\ \bibinfo {author} {\bibfnamefont {J.~K.}\ \bibnamefont {Jain}},\ }\bibfield  {title} {\bibinfo {title} {{Kohn-Sham theory of the fractional quantum Hall effect}},\ }\href {https://doi.org/10.1103/PhysRevLett.123.176802} {\bibfield  {journal} {\bibinfo  {journal} {Phys. Rev. Lett.}\ }\textbf {\bibinfo {volume} {123}},\ \bibinfo {pages} {176802} (\bibinfo {year} {2019})}\BibitemShut {NoStop}%
\bibitem [{\citenamefont {Hu}\ \emph {et~al.}(2021)\citenamefont {Hu}, \citenamefont {Murthy}, \citenamefont {Rao},\ and\ \citenamefont {Jain}}]{ANYONKS}%
  \BibitemOpen
  \bibfield  {author} {\bibinfo {author} {\bibfnamefont {Y.}~\bibnamefont {Hu}}, \bibinfo {author} {\bibfnamefont {G.}~\bibnamefont {Murthy}}, \bibinfo {author} {\bibfnamefont {S.}~\bibnamefont {Rao}},\ and\ \bibinfo {author} {\bibfnamefont {J.~K.}\ \bibnamefont {Jain}},\ }\bibfield  {title} {\bibinfo {title} {{Kohn-Sham density functional theory of Abelian anyons}},\ }\href {https://doi.org/10.1103/PhysRevB.103.035124} {\bibfield  {journal} {\bibinfo  {journal} {Phys. Rev. B}\ }\textbf {\bibinfo {volume} {103}},\ \bibinfo {pages} {035124} (\bibinfo {year} {2021})}\BibitemShut {NoStop}%
\end{thebibliography}

%apsrev4-2.bst 2019-01-14 (MD) hand-edited version of apsrev4-1.bst
%Control: key (0)
%Control: author (8) initials jnrlst
%Control: editor formatted (1) identically to author
%Control: production of article title (0) allowed
%Control: page (0) single
%Control: year (1) truncated
%Control: production of eprint (0) enabled
%

\end{document}